%% file: pp_zg_rg_plb_paper.tex
\UseRawInputEncoding
\documentclass[review,1,times]{elsarticle}
\usepackage{lineno,hyperref}
\usepackage{graphicx}
\usepackage{amsmath}
\usepackage{amssymb}
\usepackage[usenames,dvipsnames]{color}
\usepackage{rotating}
\usepackage{tabularx}

\newcommand{\pp}{p$+$p~}
\journal{Phys Lett B}

\begin{document}

\begin{frontmatter}

\title{Measurement of Groomed Jet Substructure Observables in \pp Collisions at $\sqrt{s} = 200$~GeV with STAR}
%\date{\today}

\input{star_authorlist_22June2020.tex}

\begin{abstract}
In this letter, measurements of the shared momentum fraction ($z_{\rm{g}}$) and the groomed jet radius ($R_{\rm{g}}$), as defined in the SoftDrop algorihm, are reported in \pp collisions at $\sqrt{s} = 200$ GeV collected by the STAR experiment. These substructure observables are differentially measured for jets of varying resolution parameters from $R = 0.2 - 0.6$ in the transverse momentum range $15 < p_{\rm{T, jet}} < 60$  GeV$/c$. These studies show that, in the $p_{\rm{T, jet}}$ range accessible at $\sqrt{s} = 200$ GeV and with increasing jet resolution parameter and jet transverse momentum, the $z_{\rm{g}}$ distribution asymptotically converges to the DGLAP splitting kernel for a quark radiating a gluon. The groomed jet radius measurements reflect a momentum-dependent narrowing of the jet structure for jets of a given resolution parameter, i.e., the larger the $p_{\rm{T, jet}}$, the narrower the first splitting. For the first time, these fully corrected measurements are compared to Monte Carlo generators with leading order QCD matrix elements and leading log in the parton shower, and to state-of-the-art theoretical calculations at next-to-leading-log accuracy. We observe that PYTHIA 6 with parameters tuned to reproduce RHIC measurements is able to quantitatively describe data, whereas PYTHIA 8 and HERWIG 7, tuned to reproduce LHC data, are unable to provide a simultaneous description of both $z_{\rm{g}}$ and $R_{\rm{g}}$, resulting in opportunities for fine parameter tuning of these models for \pp collisions at RHIC energies. We also find that the theoretical calculations without non-perturbative corrections are able to qualitatively describe the trend in data for jets of large resolution parameters at high $p_{\rm{T, jet}}$, but fail at small jet resolution parameters and low jet transverse momenta.  
\end{abstract}


\end{frontmatter}

%\linenumbers
\section{Introduction}
\label{sec:introduction}

Jets are well-established signals of partons, i.e. quarks and gluons, created in the high $Q^{2}$ scatterings between partons of incoming beams during high energy hadron collisions~\cite{PhysRevLett.39.1436}. These hard scattered partons, produced at high virtuality, evolve via a parton shower undergoing splitting/branching, and end in hadronization which results in a collimated stream of particles that are then clustered into jets. Jets have played a prominent role as an internal probe of partonic energy loss mechanisms in the quark-gluon plasma created in heavy-ion collisions. Refer to \cite{Connors:2017ptx} and \cite{Qin:2015srf} for recent reviews of the experimental measurements and theoretical calculations on jet quenching. An important prerequisite of such studies is the measurements of differential jet yields and jet properties related to the shower evolution and hadronization. The production of hard scattered partons is governed by $2\rightarrow2$ quantum chromodynamics (QCD) scattering at leading order (LO) and $2\rightarrow3$ at next-to-leading order (NLO). These production cross-sections for quarks and gluons can be calculated from convolutions of QCD matrix elements and Parton Distribution Functions (PDFs)~\cite{Dittmar:941455}, which are extracted using fits to experimental measurements, including but not limited to jet cross-sections at various kinematics. Given a hard scattered parton, the Dokshitzer-Gribov-Lipatov-Altarelli-Parisi (DGLAP) splitting kernels~\cite{Gribov:1972ri, Dokshitzer:1977sg, Altarelli:1977zs} describe its evolution and fragmentation based on perturbative quantum chromodynamics (pQCD). At LO, the splitting probabilities of a parton in vacuum depend on the momentum fraction of the radiated gluon and the corresponding angle of emission. 
Due to the double logarithmic structure of the splitting kernels and color coherence in QCD, the evolution is expected to follow an angular or virtuality ordered shower. Such an ordering implies that the earliest splittings are wide in angle and harder (referring to a high momentum radiated gluon). Collinear softer splittings on the other hand take place later during parton shower evolution. Therefore, the splitting probability can be described by two observables: the split's momentum fraction and its angle with respect to the parton direction. The primary focus of this letter is to study QCD and parton evolution in \pp collisions at RHIC. We establish a quantitative description of jet substructure that can serve as a reference for comparison to similar measurements in heavy-ion collisions where jet properties are expected to be modified due to jet quenching effects. 

In this letter, we present fully corrected measurements of the SoftDrop~\cite{Dasgupta:2013ihk, Larkoski:2014wba, Larkoski:2015lea} groomed momentum fraction ($z_{\rm{g}}$) and the groomed jet radius ($R_{\rm{g}}$) in \pp collisions at center-of-mass energy $\sqrt{s} = 200$ GeV. In vacuum, these measurements offer a correspondence to the DGLAP splitting functions during parton shower evolution. These observables are related to the modified mass drop tagger or SoftDrop grooming algorithm, used to remove soft, wide-angle radiation from sequentially clustered jets. This is achieved by recursively de-clustering the jet's angular-ordered branching history via the Cambridge/Aachen (C/A) clustering algorithm~\cite{Dokshitzer:1997in, Wobisch:1998wt}, which sequentially combines nearest constituents, i.e., those located closest in angle. Subjets are discarded until the transverse momenta, $p_{\rm{T,1}}$ and $p_{\rm{T,2}}$, of the subjets from the current splitting fulfill the SoftDrop condition, $
  z_{\rm{g}} = \frac{\min(p_{\rm{T,1}},p_{\rm{T,2}}) }{ p_{\rm{T,1}}+p_{\rm{T,2}}} > z_{\text{cut}}\left(\frac{R_{\rm{g}}}{R}\right)^\beta $,
where $R_{\rm{g}}$ is the groomed jet radius, the distance defined in pseudorapidity-azimuthal angle ($\eta-\phi$) space between the two surviving subjets and $R$ is the jet resolution parameter. This analysis uses $\beta=0$ and a momentum fraction cut of  $z_{\text{cut}}=0.1$~\cite{Larkoski:2014wba} to determine if a subjet at a given clustering step survives the grooming procedure. The  $z_{\rm{cut}}$ parameter is introduced to reduce sensitivity to non-perturbative effects arising from the underlying event and hadronization~\cite{Larkoski:2014wba, Larkoski:2017bvj}. It has been shown that for such a choice of $z_{\rm{cut}}$ and $\beta$, along with the usage of the C/A algorithm for de-clustering, the distribution of the resulting $z_{\rm{g}}$ converges to the vacuum splitting probability for $z>z_{\text{cut}}$ in a ``Sudakov-safe'' manner~\cite{Larkoski:2015lea}, i.e., independent of the strong coupling constant ($\alpha_s$) in the ultraviolet (UV) limit and under the fixed coupling approximation. Since the splitting kernels are defined to be independent of the momenta of initial partons, the UV limit corresponds to a jet of infinite momentum. 

The SoftDrop $z_{\rm{g}}$ was first measured by the CMS collaboration in \pp and Pb$+$Pb collisions at $\sqrt{s_{\rm{NN}}} = 5.02$ TeV at the LHC for jets with $p_{\rm{T, jet}} > 140$ GeV/$c$~\cite{Sirunyan:2017bsd}. As the measurements are not corrected for smearing due to detector effects and resolution in Pb$+$Pb, results from Monte Carlo (MC) generators, such as PYTHIA 6~\cite{Sjostrand:2006za}, PYTHIA 8~\cite{Sjostrand:2014zea} and HERWIG++~\cite{Bahr:2008pv, Bellm:2015jjp}, are convoluted with detector effects to make meaningful comparisons. Due to the granularity of the CMS hadronic calorimeter, a $R_{\rm{g}}> 0.1$ threshold was enforced which consequently introduced a bias towards wider jets in the study~\cite{Milhano:2017nzm}. 
A recent measurement from ATLAS~\cite{Aad:2019vyi} proceeded to fully unfold the SoftDrop observables for both track and calorimetric jets. It was shown that event generators, with parameters tuned to LHC data, generally reproduce the trend in p$+$p collisions, but, neither PYTHIA 8 nor HERWIG 7 were able to quantitatively describe the measurements within systematic uncertainties. Jets produced in large center-of-mass energy and high luminosity collisions at the LHC have increased sensitivity to multi-parton interactions and pileup, as compared to those at RHIC. On the other hand, due to their large jet $p_{\rm{T}}$, the measurements have typically small hadronization corrections and higher-order power corrections~\cite{Frye:2016aiz, Liu:2018ktv} due to a small $\alpha_{\rm{s}}$. 

The p$+$p collisions at RHIC provide a complementary environment to study jet structure and parton evolution. Due to the reduced center-of-mass energy (200 GeV as compared to 5.02 or 13 TeV), the study offers further insights regarding jet evolution by exploring different contributions of NLO effects and hadronization. Jets in the transverse momentum range accessible at RHIC energies are more susceptible to non-perturbative effects such as hadronization effects by virtue of their lower momenta. Some of these effects are mitigated by the SoftDrop grooming procedure~\cite{Frye:2016aiz}. In comparing jets at similar kinematics between RHIC and the LHC, it is important to note the significant difference in the quark vs.~gluon fractions with the former biased towards quark jets and the latter towards gluon jets, respectively. 

Jets used in this analysis are minimally biased since no additional selections are applied to the angular threshold. The measurements are fully corrected for detector response via a two-dimensional unfolding procedure. Thus in this letter, for the first time we present fully corrected jet substructure measurements at RHIC that are complementary to LHC measurements. Additionally, they serve as a crucial baseline for tuning event generators, validating state-of-the-art theoretical calculations of jet functions, and for using these measurements to determine medium effects in heavy-ion collisions.  

\section{Experimental Setup and Jet Reconstruction}
\label{sec:setup}

The data analyzed in this letter were collected by the STAR experiment~\cite{star_detector} in \pp collisions at $\sqrt{s}=200$~GeV in 2012.  STAR is a cylindrical detector with multiple concentric layers of detector components, including the Time Projection Chamber (TPC)~\cite{Anderson:2003ur} and a Barrel ElectroMagnetic Calorimeter (BEMC)~\cite{Beddo:2002zx}, both of which are enclosed in a 0.5 T solenoidal magnetic field. Candidate collision vertices are reconstructed with charged particle tracks from the TPC. To minimize pileup events and to ensure uniform detector acceptance, only the highest quality primary vertex in each event is selected, and its position along the beam axis is required to fall within $|z_{\rm{vertex}}| < 30$~cm from the center of the STAR detector. 

Jet finding in this analysis utilizes both the charged particle tracks from the TPC and calorimeter towers from the BEMC. Tracks are required to have more than 52\% of possible space points measured in the TPC (up to 45), a minimum of 20 measured space points, a distance of closest approach (DCA) to the primary vertex less than 1 cm, and $|\eta| < 1$. The transverse energies ($E_{\rm{T}}$) of electrons, positrons and photons both directly produced and originating from decays of neutral hadrons, are extracted from the BEMC towers with a granularity of $0.05 \times 0.05$ in $\eta-\phi$. The BEMC covers full azimuth within $|\eta|<1$. Energies deposited by charged particles in the BEMC, including electrons and positrons, are accounted for through a $100\%$ hadronic correction, i.e.,~the transverse momenta of any charged tracks that extrapolate to a tower are subtracted from the tower $E_{\rm{T}}$. Tower energies are set to zero if they become negative after this correction. Events containing tracks with $p_{\rm{T}}>$ 30 GeV/$c$ were not considered due to the poor momentum resolution for such almost straight  (low curvature) tracks in the TPC. For consistency, events with BEMC towers above the same threshold were likewise rejected.

Events were selected  online by a BEMC trigger utilizing a patch of calorimeter towers. The BEMC is split into 18 partially overlapping patches, called Jet Patches (JP), covering $1.0 \times 1.0$ in $\phi-\eta$. To fulfill the JP requirement, the combined raw ADC counts in at least one of the patches is above a certain threshold corresponding to $\sum E_{\rm{T, Tower}}>7.3$~GeV. With these aforementioned requirements on event selection, we select and analyze about 11 million triggered events. 

Towers and charged tracks with $0.2 < E_{\rm{T}} ~ (p_{\rm{T}}) < 30.0$ GeV (GeV/$c$) are clustered into jets using the anti-$k_{\rm{T}}$ algorithm from the FastJet package~\cite{Cacciari:2008gp}. Jets are reconstructed with varying resolution parameters, $R = 0.2, 0.4$ and $0.6$, and within $|\eta^{\rm{jet}}| < 1-R$ to avoid partially reconstructed jets at the edge of the acceptance. Jets are also required to have no more than 90\% of their energies provided by the BEMC towers to ensure good quality. This requirement rejects $3.4\%$ of the reconstructed jets with the effect predominantly occurring at $p_{\rm{T, jet}} \sim 15$ GeV/$c$. The fully reconstructed jets that pass the SoftDrop criteria are then considered for the study.

\section{Detector Simulation and Unfolding}
\label{sec:detector-simulation}

\begin{figure}
  \centering
  \includegraphics[width=0.8\textwidth]{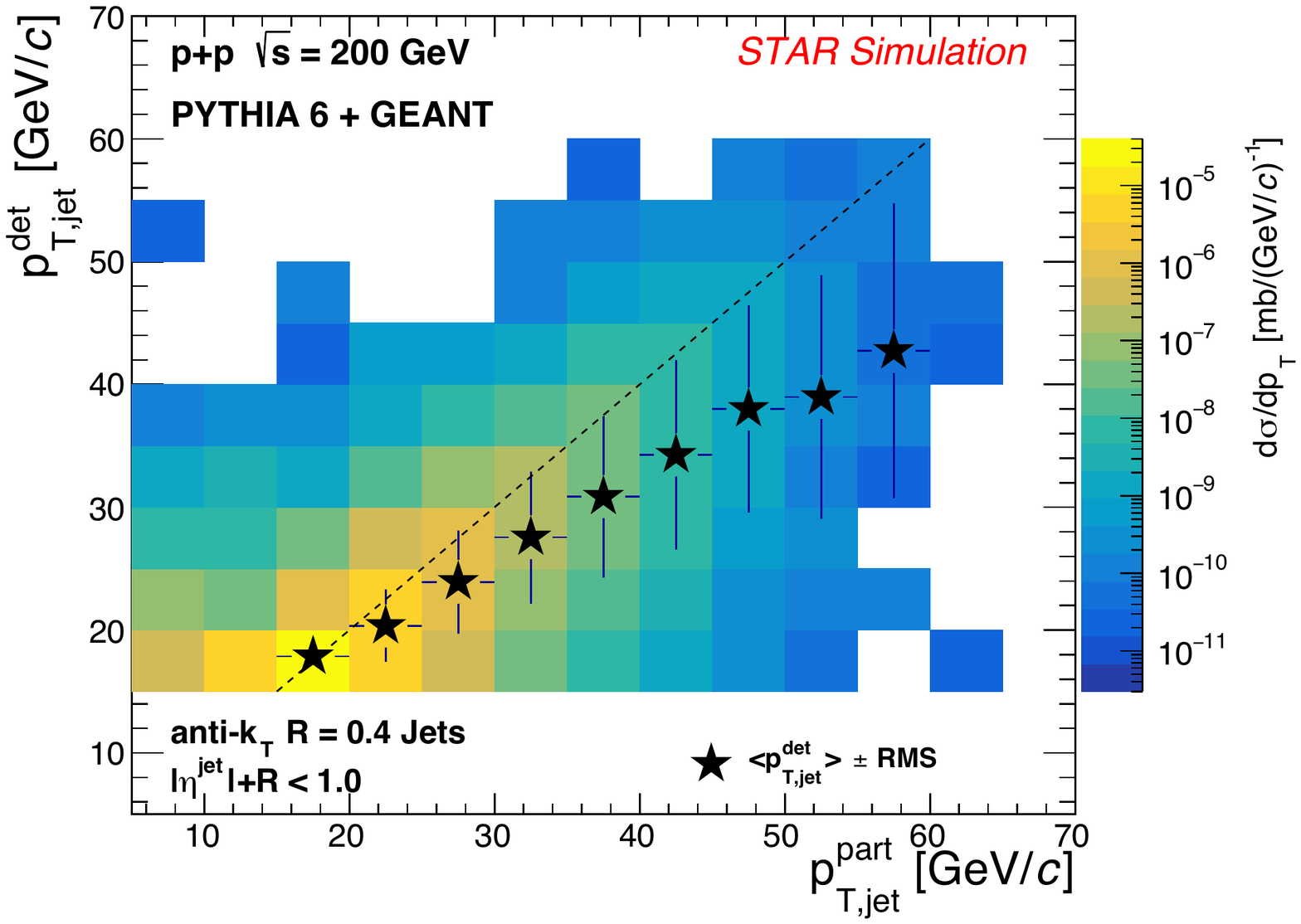}
  \caption{Detector-level jet $p_{\rm{T, jet}}^{\rm{det}}$ from PYTHIA 6 + GEANT 3 simulation for STAR detector versus PYTHIA 6 particle-level jet $p_{\rm{T, jet}}^{\rm{part}}$ for $R=0.4$ jets. The data points and the error bars represent the mean $p_{\rm{T, jet}}^{\rm{det}}$ and the width (RMS) for a given $p_{\rm{T, jet}}^{\rm{part}}$ selection.}
  \label{fig:pTResponse}
\end{figure}

\begin{figure*}
  \centering
  \includegraphics[width=.9\textwidth]{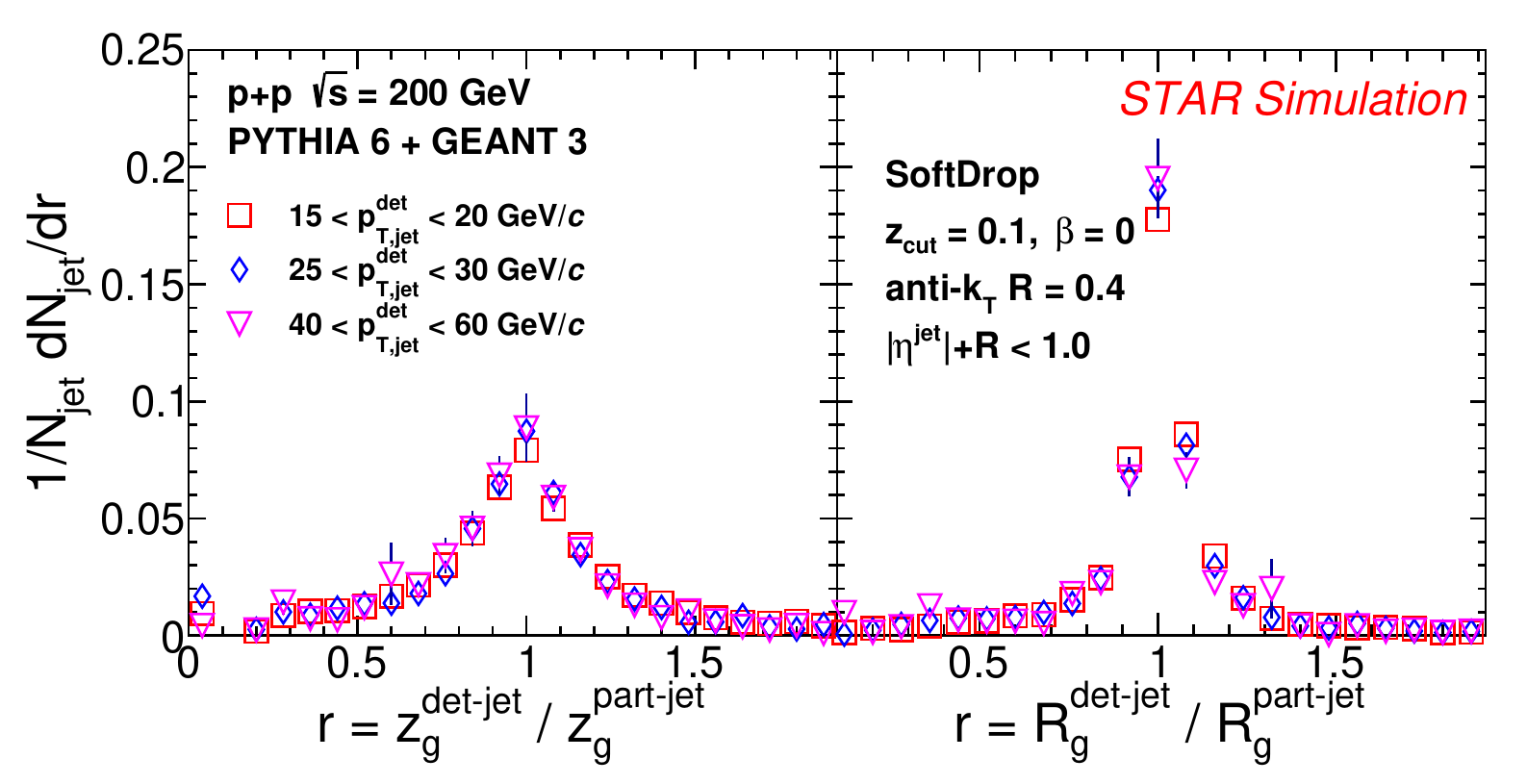}
  \caption{Detector resolutions shown as the ratio of the detector-level to the matched particle-level SoftDrop observables $z_{\rm{g}}$ (left) and $R_{\rm{g}}$ (right) for $R = 0.4$ jets with various selections of $p_{\rm{T, jet}}^{\rm{det}}$.}
  \label{fig:resolution}
\end{figure*}

In order to study the response of the STAR detector to jet substructure observables, \pp events at $\sqrt{s}=200$ GeV are generated using the PYTHIA 6.4.28~\cite{Sjostrand:2006za} event generator with the Perugia 2012 tune and CTEQ6L PDFs~\cite{Lai:1999wy}. The PYTHIA 6 version used in this analysis was further tuned to match the underlying event characteristics as measured by STAR in a recent publication~\cite{Adam:2019aml}. These generated events are then passed through a GEANT 3~\cite{Brun:1994aa} simulation of the STAR detector and embedded into zero-bias data from the same \pp run period to account for pileup contributions. For the simulated events including detector effects, identical analysis procedures including event and jet selection criteria mentioned in Sect.~\ref{sec:setup} are applied. Jets that are found from PYTHIA 6 simulations before and after the embedding procedures are hereafter referred to as particle-level and detector-level jets, respectively. Jet finding at the particle level includes weak-decaying mother particles, and their subsequent decays are simulated and the decay products are included in the detector-level jets as in real data analysis. The STAR detector response to a jet is estimated by comparing the properties of a PYTHIA 6 particle-level jet with its geometrically matched detector-level jet based on the following matching criterion, $\sqrt{(\Delta \eta)^2 + (\Delta \phi)^2} < R$, where the $\Delta$ refers to the difference between the detector- and particle-level jets in the same event and $R$ is the jet resolution parameter. With our jet quality selections, we have about 2\% of detector-level jets with $p_{\rm{T, jet}}^{\rm{det}} > 15$ GeV/$c$ that cannot be matched to particle-level jets. On the other hand, the jet finding efficiency for particle-level jets varies within 80-94\% for $15 < p_{\rm{T, jet}}^{\rm{part}} < 60$ GeV/$c$. The two dimensional $p_{\rm{T, jet}}$ response matrix for $R=0.4$ jets is shown in Fig.~\ref{fig:pTResponse}. We find that due to detector effects the mean $p_{\rm{T, jet}}^{\rm{det}}$ (shown in the black filled markers) is significantly smaller than the corresponding $p_{\rm{T, jet}}^{\rm{part}}$. For the jet substructure observables, the detector response is shown in Fig.~\ref{fig:resolution}, quantified by the ratio of detector-level jet quantity to the matched particle-level jet quantity for a variety of $p_{\rm{T, jet}}^{\rm{det}}$ selections. Cases where one of the jets (matched detector- or particle-level jet) does not pass the SoftDrop criterion are shown in the first bin on the x-axis in the left panel of Fig.~\ref{fig:resolution}. The ratios are peaked at unity and independent of $p_{\rm{T, jet}}^{\rm{det}}$, which facilitates correcting the measurements for detector effects via a two-dimensional (e.g., $p_{\rm{T, jet}}$ and $z_{\rm{g}}$) unfolding procedure. 

 \begin{figure*}
  \centering
  \includegraphics[width=.4\textwidth]{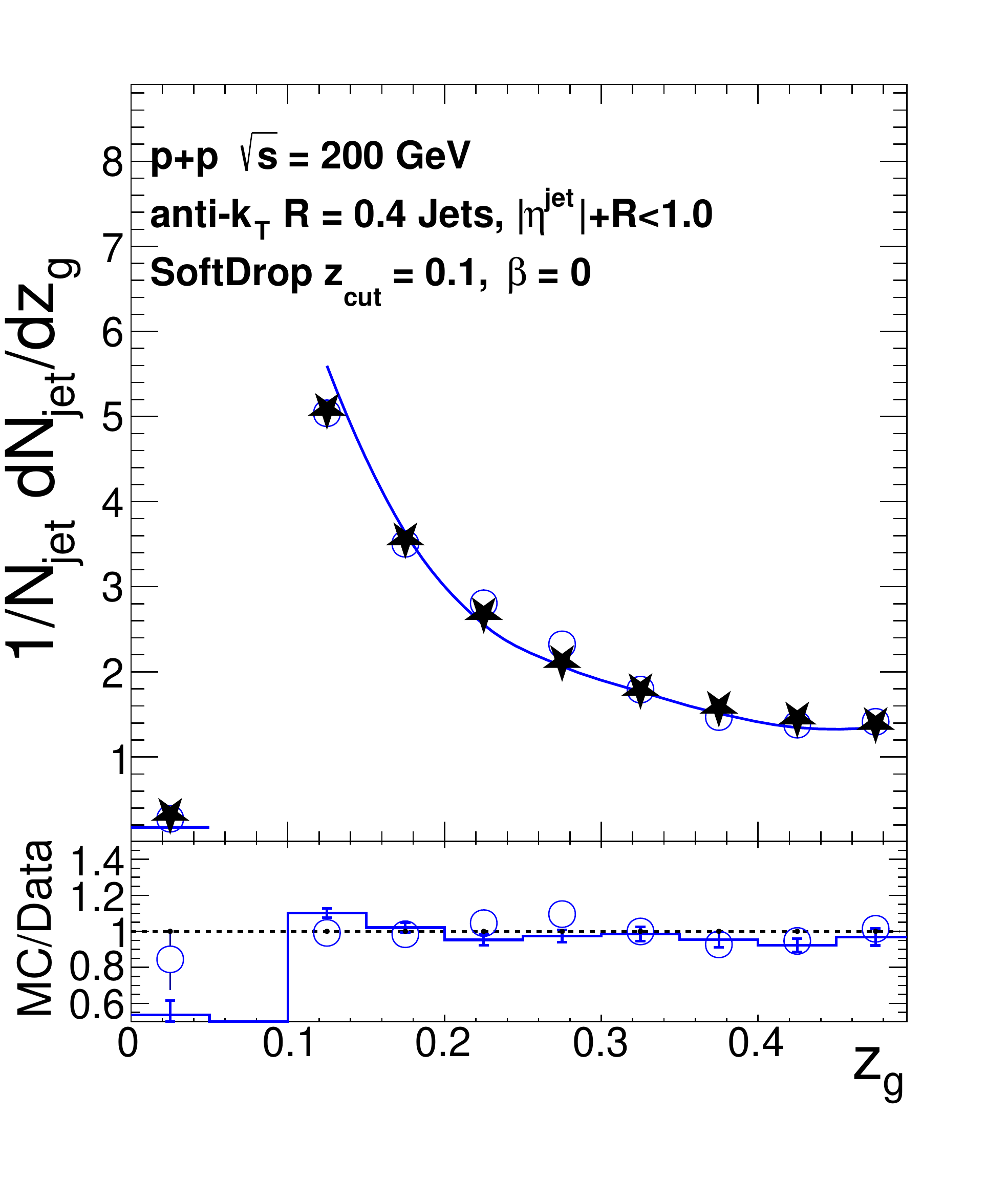}
  \includegraphics[width=.4\textwidth]{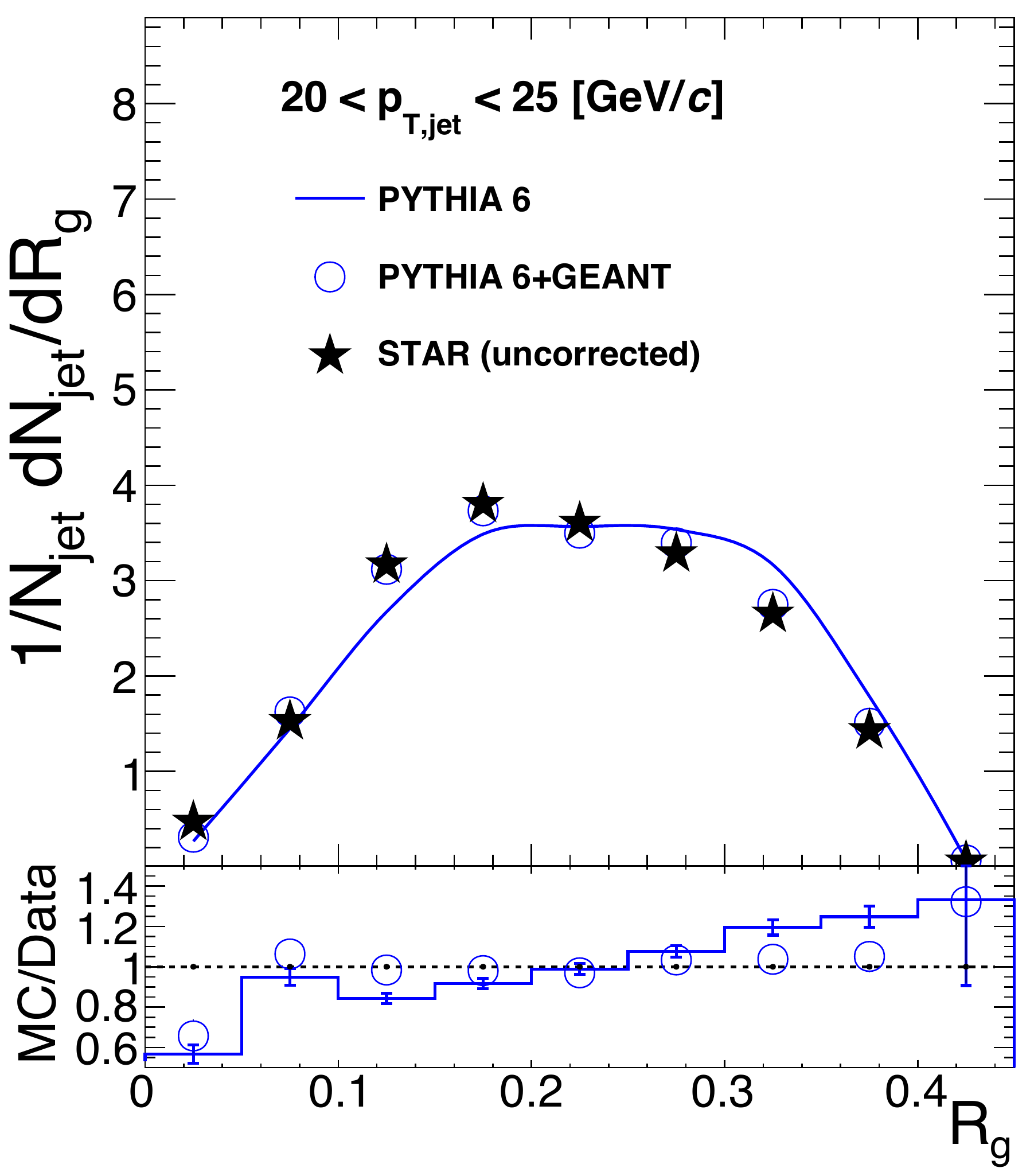}
  \caption{Comparisons of the SoftDrop $z_{\rm{g}}$ (left) and $R_{\rm{g}}$ (right) distributions in raw data to PYTHIA 6 and PYTHIA 6$+$GEANT 3 simulations. The bottom panels show the ratio of MC to raw data.}
  \label{fig:rawDataComp}
\end{figure*}

For anti-$k_{\rm{T}}$, $R=0.4$ jets with $20 < p_{\rm{T, jet}} < 25$ GeV/$c$, the tuned PYTHIA 6 (blue solid line), PYTHIA 6$+$GEANT 3 simulation (blue open circles) and uncorrected data (filled black stars) distributions are shown in Fig.~\ref{fig:rawDataComp} for $z_{\rm{g}}$ on the left and $R_{\rm{g}}$ on the right. The bottom panels show the ratio of simulation to data where we observe a good agreement between detector-level simulation and data. In comparing the particle-level and detector-level PYTHIA 6 distributions, we see small but statistically significant differences due to the detector response which we correct for via an unfolding method described below.

The SoftDrop $z_{\rm{g}}$ and $R_{\rm{g}}$ distributions in this analysis are unfolded to the particle level to correct for detector effects including smearing and bin migration. The detector response for substructure observables peaks at unity and is independent of $p_{\rm{T,jet}}$, as shown in Fig.~\ref{fig:resolution}, and the resulting four-dimensional (i.e., detector- and particle-level $p_{\rm{T, jet}}$ and $z_{\rm{g}}$ or $p_{\rm{T, jet}}$ and $R_{\rm{g}}$) response matrix is utilized in the correction procedure. Two-dimensional Bayesian unfolding~\cite{DAgostini:1994fjx} is performed to take into account non-diagonal bin-to-bin migrations both in jet $p_{\rm{T}}$ and SoftDrop observables, using the tools available in the RooUnfold package~\cite{roounfold} with four iterations as the default parameter. As a consequence of the detector simulation reproducing the uncorrected data as shown in Fig.~\ref{fig:rawDataComp} and the resolutions for the SoftDrop observables being relatively narrow and independent of $p_{\rm{T, jet}}$ as shown in Fig.~\ref{fig:resolution}, the unfolding procedure converges and is numerically stable. The priors in the unfolding procedure are taken from the PYTHIA 6 simulation and their variations are studied as a source of systematic uncertainty.

\section{Systematic uncertainties}
\label{sec:unfolding}

There are two main categories of systematic uncertainties considered in this analysis. The first is related to the reconstruction performance of the STAR detector, including the uncertainty on the tower gain calibration ($3.8\%$) and the absolute tracking efficiency ($4\%$). The other source of systematic uncertainty is due to the analysis procedure, i.e., the use of hadronic correction (as described in Sec.~\ref{sec:setup}) and the unfolding procedure. The correction to the tower energy, based on the momenta of the matched tracks, is varied by subtracting half of the momenta of the matched tracks from their corresponding tower $E_{\rm{T}}$. With regards to the unfolding procedure, the uncertainties include the variation of the iteration parameter from 2--6 with 4 as the nominal value, and a variation of the input prior shape for $z_{\rm{g}},R_{\rm{g}}$ and $p_{\rm{T}}$ individually by using PYTHIA 8 and HERWIG 7. We estimated the effect of different sources on the final results by varying the detector simulation, following the same unfolding procedure and comparing to the nominal result. Since we are reporting self-normalized distributions, the luminosity uncertainty for the given data taking period is not considered. The total systematic uncertainties for the $z_{\rm{g}}$ and $R_{\rm{g}}$ measurements, calculated by adding individual sources in quadrature, are presented in Tab.~\ref{tab:zg} and \ref{tab:rg} for $R=0.4$ jets in the range $20 < p_{\rm{T, jet}} < 25$  GeV$/c$. For both measurements, the largest systematic uncertainty results from the unfolding procedure. The total systematic uncertainties for these SoftDrop observables decrease slightly as the jet resolution parameter increases.

\begin{table*}
\begin{center}
\begin{tabular}{| c | c | c | c | c | c |}
\hline
Source $/$ & Hadronic & Tower & Tracking & Unfolding & Total \\ 
Range in $z_{\rm{g}}$ & Correction & Gain &  Efficiency &  &  \\  [0.5ex]
 \hline\hline
 [0.10, 0.15] & 0.4\% & 2\% & 1.7\% & 2.9\% & 3.9\% \\
\hline
 [0.25, 0.30] & $\approx 0\%$  & 2.3\% & 1.5\% & 5.2\% & 5.8\% \\ 
\hline
 [0.45, 0.50] & 0.6\% & 1.6\% & 1.9\% & 6.8\% & 7.3\% \\  [0.5ex]
 \hline
% \hline
\end{tabular}
\caption{\label{tab:zg}  Uncertainties on the SoftDrop $z_{\rm{g}}$ measurement for $R=0.4$ jets with $20 < p_{\rm{T, jet}} < 25$ GeV/$c$ as a representative jet sample.}
\end{center}
\end{table*}

\begin{table*}
\begin{center}
\begin{tabular}{| c | c | c | c | c | c |}
\hline
Source $/$ & Hadronic & Tower & Tracking & Unfolding & Total \\ 
Range in $R_{\rm{g}}$ & Correction & Gain &  Efficiency &  &  \\  [0.5ex]
\hline\hline
 [0.10 - 0.15] & 2\% & 2.2\% & 5.6\% & 7.6\% & 9.9\% \\ 
\hline
 [0.20 - 0.25] & 0.5\% & 1.1\% & 0.2\% & 1.9\% & 2.2\% \\ 
\hline
 [0.30 - 0.35] & 1.6\% & 2.8\% & 2.6\% & 9.1\% & 10\% \\ 
\hline
 [0.40 - 0.45] & 8.4\% & 2.7\% & 20.6\% & 40.3\% & 46.15\% \\ 
\hline
%\hline
\end{tabular}
\caption{\label{tab:rg} Uncertainties on the SoftDrop $R_{\rm{g}}$ measurement for $R=0.4$ jets with $20 < p_{\rm{T, jet}} < 25$ GeV/$c$ as a representative jet sample.}
\end{center}
\end{table*}

\section{Results}

The fully corrected $z_{\rm{g}}$ and $R_{\rm{g}}$ measurements are compared to leading order event generators, PYTHIA 6, PYTHIA 8 and HERWIG 7. Since PYTHIA 6 events include weak-decaying mother particles at the particle level, we generate PYTHIA 8 and HERWIG 7 events with the same requirement. We note that for the observables discussed in this letter, we do not observe a significant difference between including these mother particles or their decay daughters.
The parton shower implementations are different amongst the models, with PYTHIA 6 and PYTHIA 8 featuring virtuality ordered shower in contrast to HERWIG 7 with angular ordering. The showers in all three models are however leading-log with all order shower expansion. The description of the underlying event in PYTHIA 6 is based on the Perugia 2012 tune~\cite{PhysRevD.82.074018} and further tuned to match data from RHIC, whereas PYTHIA 8 uses the Monash 2013 tune which was based on the LHC data~\cite{Skands:2014pea}. The HERWIG 7 calculations use the EE4C underlying event tune~\cite{Seymour:2013qka}.

\begin{figure*}
  \centering
  \includegraphics[width=.95\textwidth]{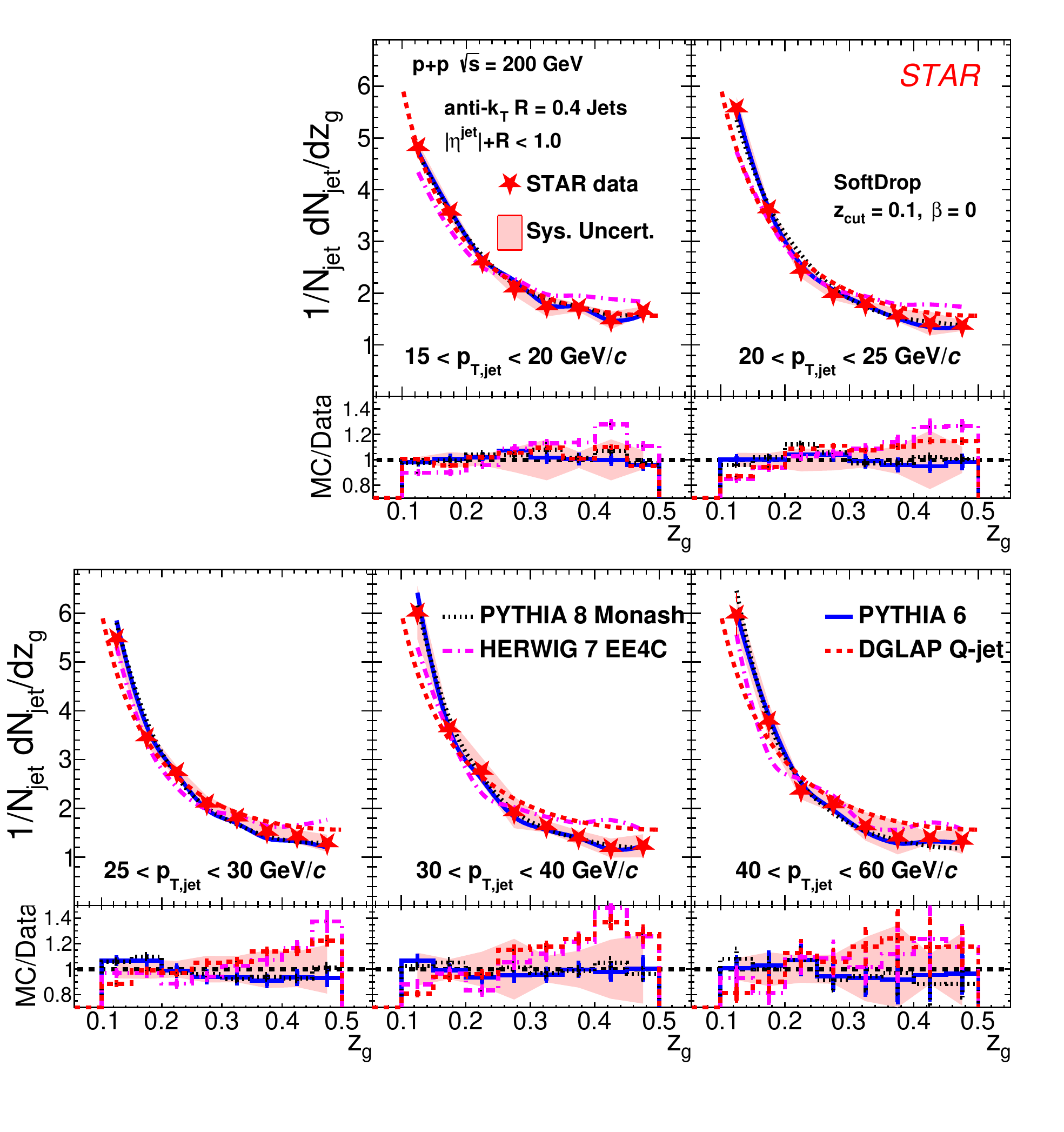}
  \caption{Distribution of the SoftDrop $z_{\rm{g}}$ in \pp collisions at $\sqrt{s} = 200$ GeV for anti-k$_{\rm{T}}~R=0.4$ jets of varying transverse momenta ($15 < p_{\rm{T, jet}} < 20$ GeV/$c$ in top middle to $40 < p_{\rm{T, jet}} < 60$ GeV/$c$ in bottom right). The data are presented as red stars and the systematic uncertainties as shaded red regions (statistical errors are in most cases smaller than the marker size). The measurements are compared to PYTHIA 8 (Monash 2013 Tune, dotted black line), PYTHIA 6 (Perugia 2012 Tune, solid blue line), and HERWIG 7 (EE4C Tune, dash-dotted magenta line). The data are also compared to the DGLAP splitting kernel for quark jets in all the panels shown in red dashed line. The corresponding bottom panels show the ratio of MC to the fully corrected data.}
  \label{fig:zgR4}
\end{figure*}

The fully corrected $z_{\rm{g}}$ measurements for jets of varying $p_{\rm{T, jet}}$ are compared to MC predictions as shown in Fig.~\ref{fig:zgR4}. In addition, we show the DGLAP splitting function at leading order for a quark emitting a gluon, with the functional form $\left(0.313 \frac{1+z^{2}}{1-z} + \frac{1 + (1-z)^{2}}{z} \right)$ as the red dashed lines where $z$ is defined as the radiated object's energy fraction with respect to the original parton. The different panels present results for jets varying from low $p_{\rm{T, jet}}$ in the top middle to high $p_{\rm{T, jet}}$ in the bottom right. We observe a more symmetric splitting function (larger mean $z_{\rm{g}}$ or, consequently, a flatter shape) at lower $p_{\rm{T, jet}}$ that gradually tends towards a more asymmetric function (smaller mean $z_{\rm{g}}$) at higher $p_{\rm{T, jet}}$. The measurements also indicate a $p_{\rm{T, jet}}$-independent $z_{\rm{g}}$ shape slightly steeper than the theoretical limit around $p_{\rm{T, jet}} > 30$ GeV/$c$ within our kinematic range. With symmetric splitting functions, the probability to radiate a high-$z$ gluon is enhanced as opposed to an asymmetric splitting function dominated by low-$z$ emissions. This evolution from a symmetric to asymmetric splitting function with increasing $p_{\rm{T, jet}}$ is consistent with the pQCD expectation that a high-momentum parton has an enhanced probability to radiate a soft gluon. 
Such behavior is captured by both angular and virtuality ordered parton shower models. With default hadronization turned on, PYTHIA 6, PYTHIA 8 and HERWIG 7 describe the qualitative shape as observed in these measurements. To compare more quantitatively, the bottom panels show the ratio of the model calculations to data, and the shaded red region represents the total systematic uncertainty in data. Both PYTHIA versions are able to describe the $z_{\rm{g}}$ measurements. However, HERWIG 7 seems to prefer more symmetric splittings, specially for the highest $p_{\rm{T, jet}}$ ranges. 

\begin{figure*}
  \centering
  \includegraphics[width=.95\textwidth]{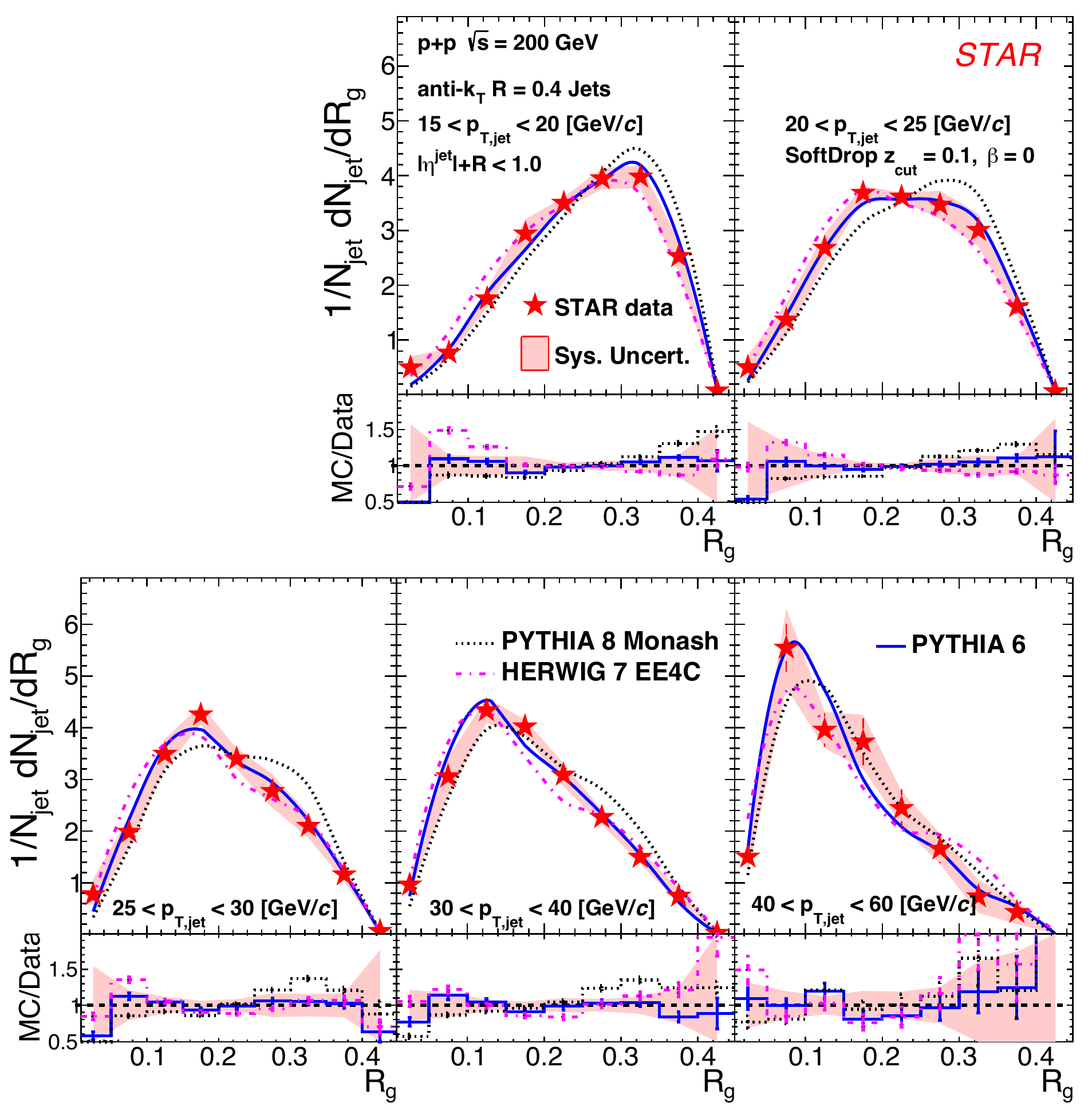}
  \caption{Measurement of the SoftDrop $R_{\rm{g}}$ in \pp collisions at $\sqrt{s} = 200$ GeV for anti-k$_{\rm{T}}~R=0.4$ jets. The description of the panels, symbols and lines is the same as for Fig.~\ref{fig:zgR4}.}
  \label{fig:rgR4}
\end{figure*}

The SoftDrop $R_{\rm{g}}$ distributions for $R=0.4$ jets are presented in Fig.~\ref{fig:rgR4}. They show a momentum-dependent narrowing of the jet structure as reflected in a shift to smaller values as the jet transverse momentum increases. The measured  $R_{\rm{g}}$ distributions are qualitatively reproduced by all event generators. HERWIG 7 shows a slight tendency towards smaller $R_{\rm{g}}$, while PYTHIA 8 prefers a systematically wider $R_{\rm{g}}$ distribution. For $R=0.4$ jets, PYTHIA 6 is able to quantitatively describe data, whilst neither PYTHIA 8 nor HERWIG 7 is able to explain both $z_{\rm{g}}$ and $R_{\rm{g}}$ observables simultaneously within the experimental systematic uncertainties.

\begin{figure*}
  \centering
  \includegraphics[width=.89\textwidth]{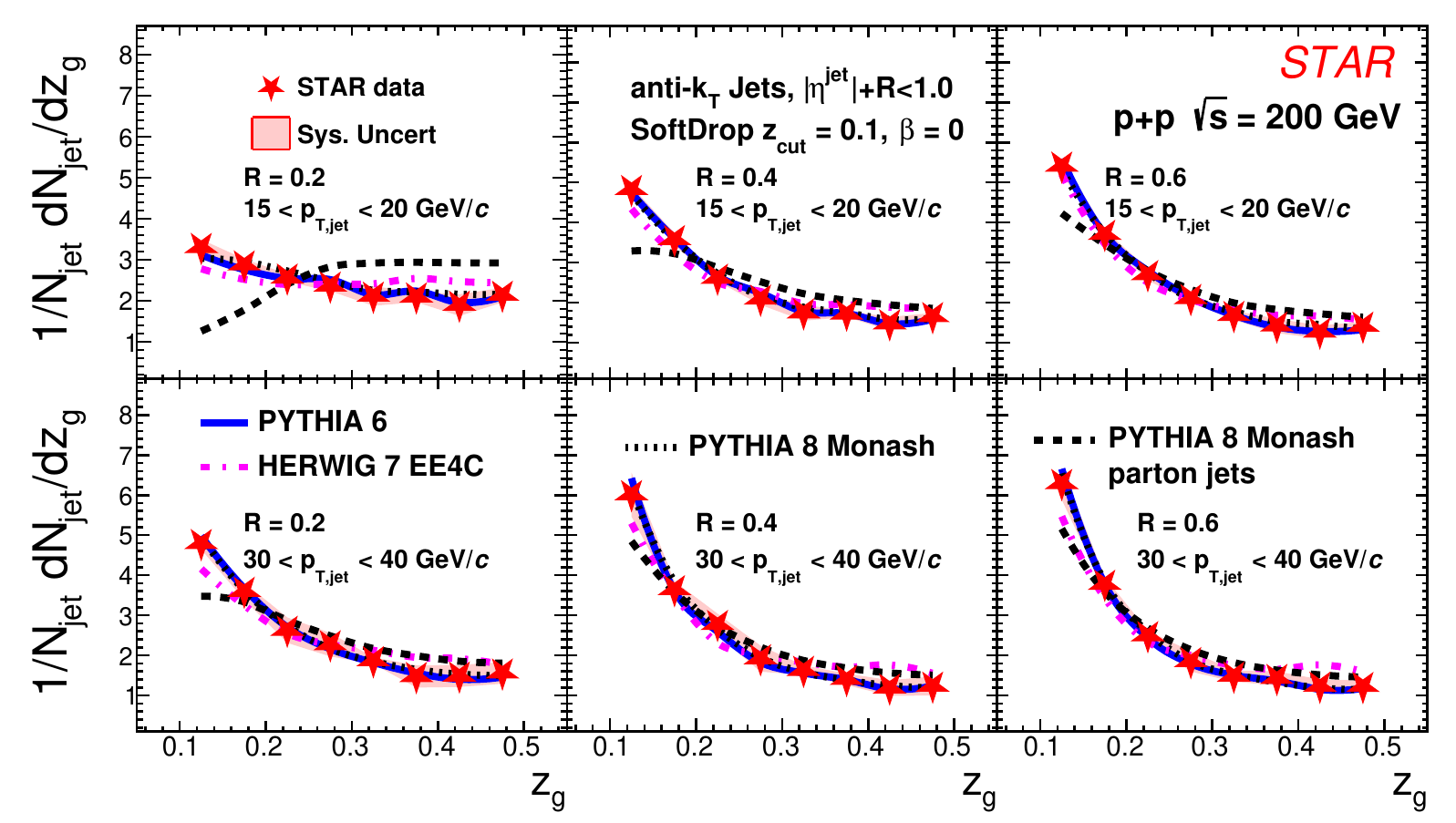}
  \caption{Radial scans of the SoftDrop $z_{\rm{g}}$ in \pp collisions at $\sqrt{s} = 200$ GeV for anti-k$_{\rm{T}}~R=0.2$ (left), $R=0.4$ (middle) and $R=0.6$ (right) jets of varying transverse momenta ($15 < p_{\rm{T, jet}}< 20$ GeV/$c$ and  $30 < p_{\rm{T, jet}} < 40$ GeV/$c$ in the top and bottom rows respectively). The descriptions of the symbols and lines is the same as for Fig.~\ref{fig:zgR4}. The data are also compared to PYTHIA 8 parton jets without hadronization shown as black dashed lines.}
  \label{fig:rScanZg}
\end{figure*}

\begin{figure*}
  \centering
  \includegraphics[width=.89\textwidth]{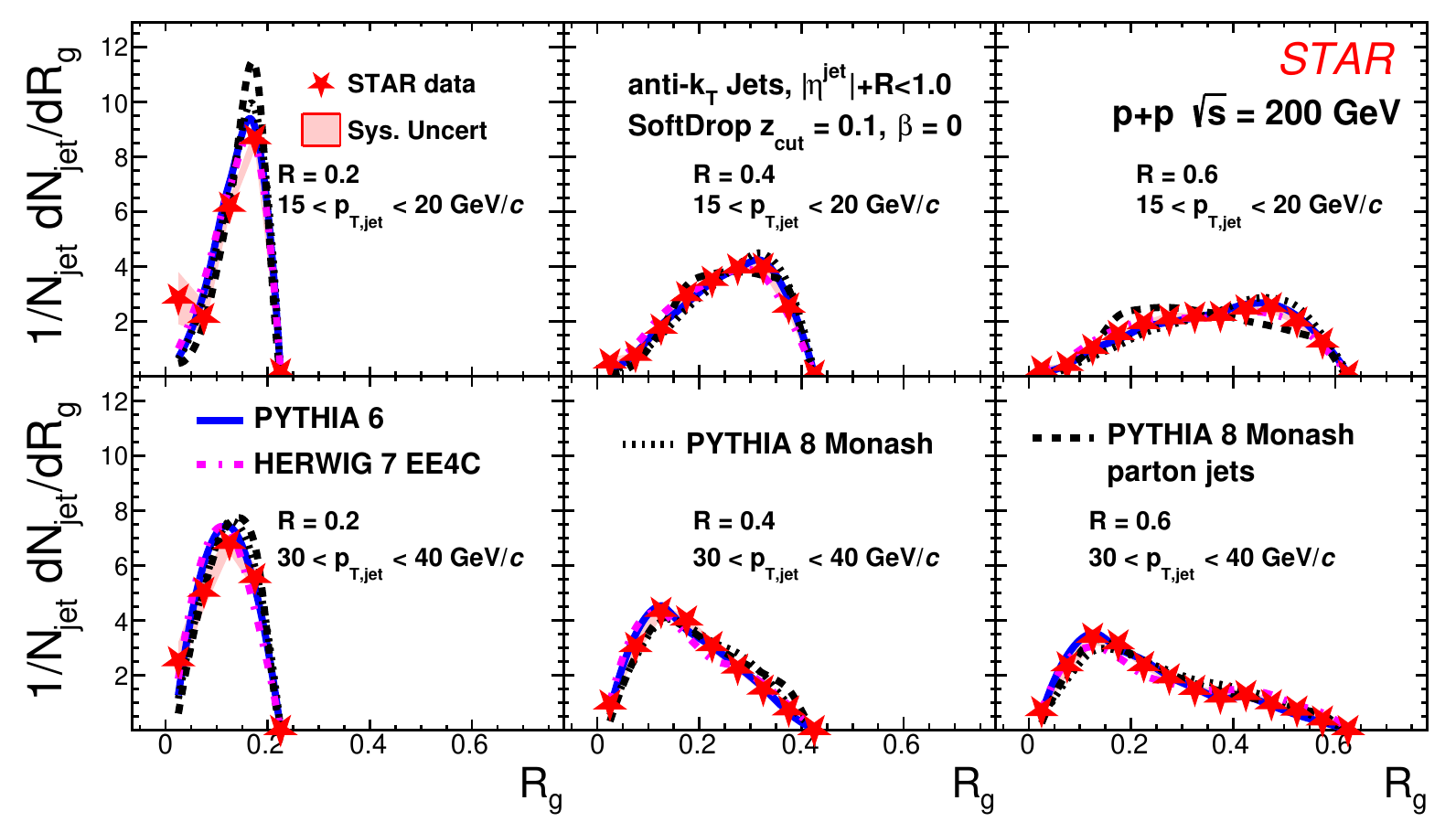}
  \caption{Radial scans of the SoftDrop $R_{\rm{g}}$ in \pp collisions at $\sqrt{s} = 200$ GeV. The different panels and calculations are similar as described in Fig.~\ref{fig:rScanZg}.}
  \label{fig:rScanRg}
\end{figure*}

Figures~\ref{fig:rScanZg} and ~\ref{fig:rScanRg} show, respectively, the measurements of $z_{\rm{g}}$ and $R_{\rm{g}}$ for varying jet resolution parameter ($R = 0.2, 0.4, 0.6$). The top row is for jets with $15 < p_{\rm{T, jet}} < 20$ GeV/$c$ and the bottom row for jets with $30 < p_{\rm{T, jet}} < 40$ GeV/$c$. Jets with smaller resolution parameters and lower $p_{\rm{T, jet}}$ display stronger $z_{\rm{g}}$ shape modifications with respect to the ideal DGLAP splitting function, and do not reproduce the characteristic $1/z$ shape seen at higher $p_{\rm{T, jet}}$. The narrowing of the $R_{\rm{g}}$ with increasing $p_{\rm{T, jet}}$ becomes more significant for jets of larger resolution parameters. The flattening of the $z_{\rm{g}}$ shape for jets with $R=0.2$ and low $p_{\rm{T, jet}}$ are due to stringent kinematic constraints on the phase space available. This interpretation is further substantiated by the observation that the $R_{\rm{g}}$ distribution is narrowing with decreasing $R$ as seen in  Fig.~\ref{fig:rScanRg}, which is a direct consequence of virtuality/angular ordering and decreasing jet finding radius. The dashed black curve shows the $z_{\rm{g}}$ and $R_{\rm{g}}$ distributions from PYTHIA 8 without hadronization (parton jets). We find that hadronization, as described in PYTHIA 8, tends to create softer $z_{\rm{g}}$ or more asymmetric splittings, but has very little impact on the $R_{\rm{g}}$ observable. 

\begin{figure*}
  \centering
  \includegraphics[width=.49\textwidth]{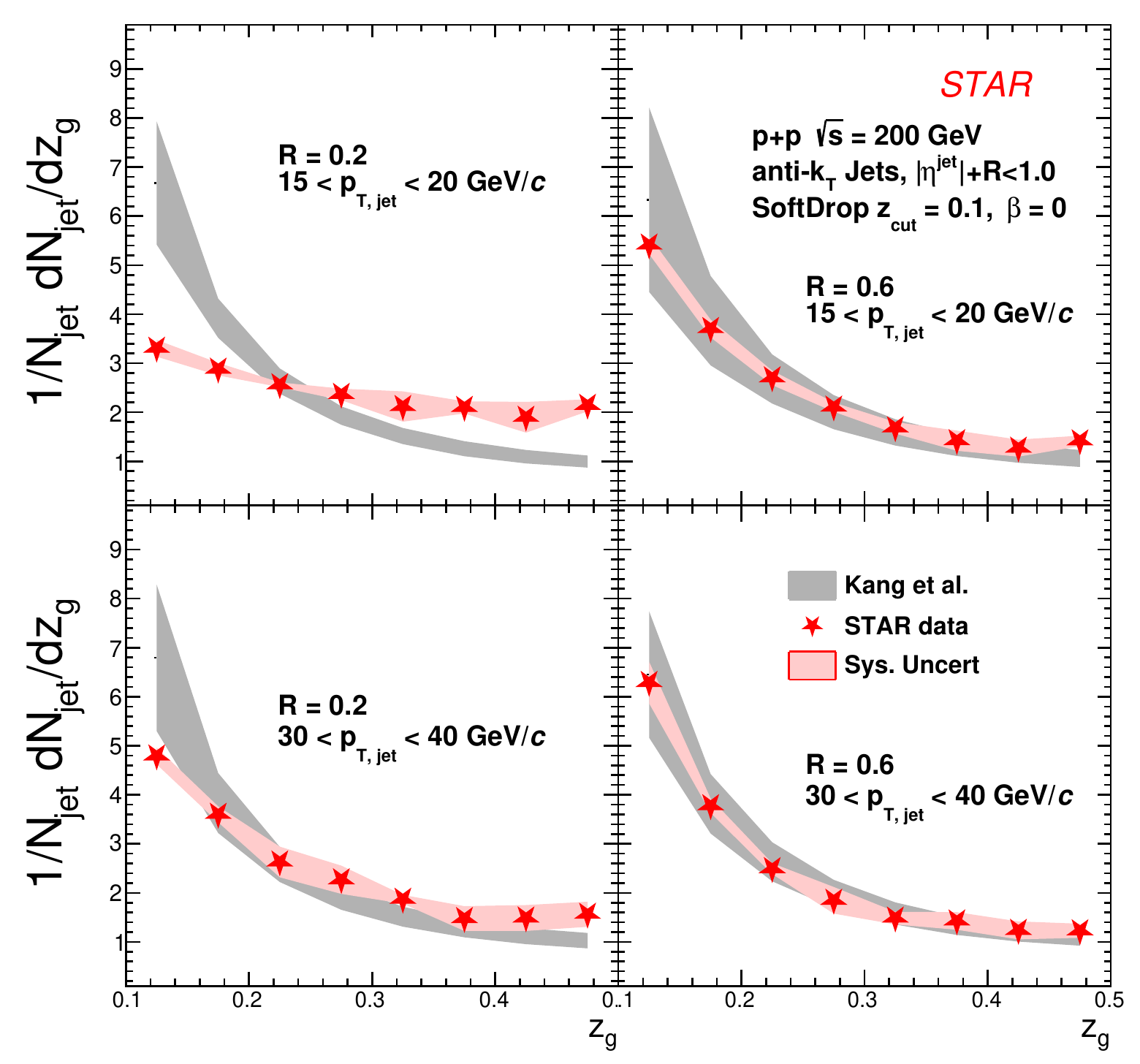}
  \includegraphics[width=.49\textwidth]{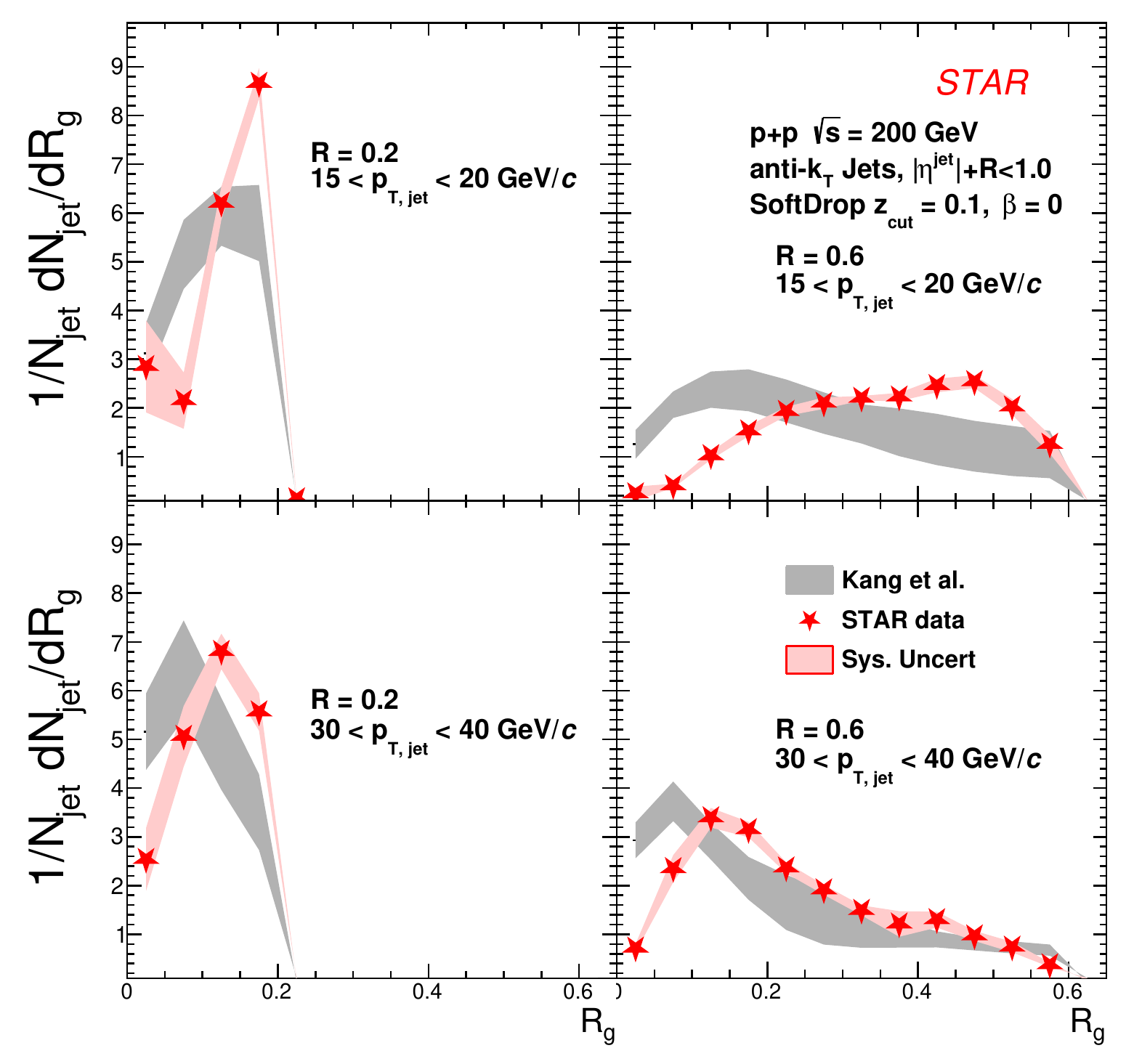}
  \caption{Comparisons of fully corrected STAR data (red markers with red shaded area as systematic uncertainties) for $z_{\rm{g}}$ (left panels) and $R_{\rm{g}}$ (right panels) with theoretical calculations at next-to-leading-log accuracy at the parton level shown as gray shaded bands. The top and bottom panels show comparisons for $15 < p_{\rm{T, jet}}< 20$ GeV/$c$ and $30 < p_{\rm{T, jet}} < 40$ GeV/$c$ respectively. In each of the 4-panel plots, the left and right columns are for jets of $R=0.2$ and $R=0.6$.}
  \label{fig:ThCompZgRg}
\end{figure*}

Due to recent advances in theoretical calculations regarding jets of small resolution parameters and low momenta~\cite{Kang:2019prh, Tripathee:2017ybi}, we can now compare our fully corrected data to predictions at next-to-leading-log accuracy in Fig.~\ref{fig:ThCompZgRg} for $z_{\rm{g}}$ (left panels) and $R_{\rm{g}}$ (right panels). The systematic uncertainty in the theoretical calculations (gray shaded band) arises from QCD scale variations, including the $p_{\rm{T}}$-hard scale, the jet scale ($p_{\rm{\rm{T, jet}}} \cdot R$) and the scales associated with the substructure observables mentioned here~\cite{Kang:2019prh}. We note that the systematic uncertainties for the calculations are large for the kinematic range studied in this measurement. These predictions are for jets at the parton level without non-perturbative corrections. The calculations for $z_{\rm{g}}$ significantly deviate from data for jets of smaller resolution parameters and lower $p_{\rm{T}}$, with the agreement getting better as the jet $R$ and $p_{\rm{T}}$ increase. On the other hand, the predictions for the $R_{\rm{g}}$ show large discrepancies with data for all of the jet resolution parameters and momenta except for the largest resolution parameter and highest $p_{\rm{T, jet}}$ where the shape gets closer to the data. These comparisons highlight the need for more realistic calculations, including corrections arising from non-perturbative effects and higher-order corrections to further quantitatively understand the jet substructure.  

\section{Summary}
\label{sec:ppsummary}

In summary, we presented the first fully corrected SoftDrop $z_{\rm{g}}$ and $R_{\rm{g}}$ measurements for inclusive jets produced in \pp collisions at $\sqrt{s} =  200$ GeV of varying resolution parameters in the range $15 < p_{\rm{T, jet}} < 60$ GeV/$c$ . The $z_{\rm{g}}$ distribution converges towards an approximately $p_{\rm{T, jet}}$-independent shape above 30 GeV/$c$ which is slightly more asymmetric than the ideal DGLAP splitting function. On the other hand, the $R_{\rm{g}}$ distribution shows a narrowing with increasing $p_{\rm{T, jet}}$. We observe that at lower transverse momenta, jets are more likely to have a wider substructure with more symmetric splitting within the jet. The RHIC-tuned PYTHIA 6 is able to reproduce both jet substructure observables, while PYTHIA 8 and HERWIG 7 are unable to simultaneously describe both scales of the jet evolution. The impact of the hadronization process is investigated using PYTHIA 8. We note that at small jet resolution parameters and low $p_{\rm{T, jet}}$, the $z_{\rm{g}}$ is sensitive to hadronization effects resulting in a significant enhancement of asymmetric splitting, whereas for larger resolution parameters, $0.4$ and $0.6$, the effect is moderate and only results in a minor change towards more asymmetric splitting. On the other hand, the SoftDrop $R_{\rm{g}}$ is observed to be less sensitive to hadronization. We also showed comparisons to theoretical calculations at jet scales closer to the fundamental QCD scale, i.e.,~for jets with small momenta. Such comparisons to data highlight the need for continued theoretical studies into the exact interplay between measured hadronic jet substructure observables and the underlying partonic splitting at RHIC energies. These studies offer a unique opportunity to further tune MC event generators and for understanding higher order effects on jet evolution at RHIC kinematics.

\section*{Acknowledgement}
We thank Jesse Thaler, Yacine Mehtar-Tani, Felix Ringer and Zhangbo Kang for useful discussions on the topic of jet substructure and SoftDrop.  
We thank the RHIC Operations Group and RCF at BNL, the NERSC Center at LBNL, and the Open Science Grid consortium for providing resources and support.  This work was supported in part by the Office of Nuclear Physics within the U.S. DOE Office of Science, the U.S. National Science Foundation, the Ministry of Education and Science of the Russian Federation, National Natural Science Foundation of China, Chinese Academy of Science, the Ministry of Science and Technology of China and the Chinese Ministry of Education, the National Research Foundation of Korea, Czech Science Foundation and Ministry of Education, Youth and Sports of the Czech Republic, Hungarian National Research, Development and Innovation Office, New National Excellency Programme of the Hungarian Ministry of Human Capacities, Department of Atomic Energy and Department of Science and Technology of the Government of India, the National Science Centre of Poland, the Ministry  of Science, Education and Sports of the Republic of Croatia, RosAtom of Russia and German Bundesministerium fur Bildung, Wissenschaft, Forschung and Technologie (BMBF) and the Helmholtz Association.

%% References with BibTeX database:
\bibliographystyle{elsarticle-num}
\bibliography{pp_zg_rg_plb_paper}

\end{document}

%% file: star_authorlist_22June2020.tex
\author{
J.~Adam$^{6}$,
L.~Adamczyk$^{2}$,
J.~R.~Adams$^{39}$,
J.~K.~Adkins$^{30}$,
G.~Agakishiev$^{28}$,
M.~M.~Aggarwal$^{41}$,
Z.~Ahammed$^{61}$,
I.~Alekseev$^{3,35}$,
D.~M.~Anderson$^{55}$,
A.~Aparin$^{28}$
% ...
}

% ...
\address{$^{2}$AGH University of Science and Technology, FPACS, Cracow 30-059, Poland}
\address{$^{3}$Alikhanov Institute for Theoretical and Experimental Physics NRC "Kurchatov Institute", Moscow 117218, Russia}
% ...
\address{$^{6}$Brookhaven National Laboratory, Upton, New York 11973}
% ...
\address{$^{28}$Joint Institute for Nuclear Research, Dubna 141 980, Russia}
% ...
\address{$^{30}$University of Kentucky, Lexington, Kentucky 40506-0055}
% ...
\address{$^{35}$National Research Nuclear University MEPhI, Moscow 115409, Russia}
% ...
\address{$^{39}$Ohio State University, Columbus, Ohio 43210}
% ...
\address{$^{41}$Panjab University, Chandigarh 160014, India}
% ...
\address{$^{55}$Texas A\&M University, College Station, Texas 77843}
% ...
\address{$^{61}$Variable Energy Cyclotron Centre, Kolkata 700064, India}
% ...
